\documentclass[usenatbib,usegraphicx,useAMS]{mn2e}
\usepackage{times}
  
\begin{document}


\title{VLA observations of 6-cm excited OH}

\author[Palmer, et al.]
{Patrick Palmer,$^1$\thanks{E-mail: ppalmer@oskar.uchicago.edu}
W. M. Goss,$^2$\thanks{E-mail: mgoss@nrao.edu}  
and J. B. Whiteoak$^3$\thanks{E-mail: John.Whiteoak@atnf.csiro.au} \\
$^1$Department of Astronomy and Astrophysics, University of
Chicago, 5640 S. Ellis Ave., Chicago, IL 60637 \\
$^2$National Radio Astronomy Observatory, P. O. Box O, Socorro, NM 87801 \\
$^3$Australia Telescope National Facility, CSIRO, PO Box 76, Epping, NSW2121, Australia 
\\
}

\maketitle
\begin{abstract}
The VLA was used to determine precise positions for 4765-MHz OH maser
emission sources toward star-forming regions which had been observed about
seven months earlier with the Effelsberg 100-meter telescope.  The
observations were successful for K3-50, DR21EX, W75N, and W49A.  No line
was detected toward S255: this line had decreased to less than 5 per cent of
the flux density observed only seven months earlier.  The time-variability
of the observed features during the past 30 years is summarised.  In
addition, to compare with the Effelsberg observations, the 4750-MHz and
4660-MHz lines were observed in W49A.  These lines were found to originate
primarily from an extended region which is distinguished as an exceptional
collection of compact continuum components as well as by being the
dynamical centre of the very powerful H$_2$O outflow.
\end{abstract}
 

\section{Introduction}

The low lying $^2\Pi_{1/2}, J=1/2$ rotationally excited state of OH has
three hyperfine transitions in the 6-cm band: the F=0$\rightarrow$1 at
4660.242 MHz, the F=1$\rightarrow$1 at 4750.656 MHz, and the
F=1$\rightarrow$0 at 4765.562 MHz (Lovas, Johnson, \& Snyder 1979).  We
have previously reported observations conducted with the Very Large Array
(VLA) of the National Radio Astronomy Observatory\footnotemark[1]
\footnotetext[1]{The National Radio Astronomy Observatory is a
facility of the National Science Foundation operated under cooperative
agreement by Associated Universities, Inc.} of these lines in three
sources: W3 (Gardner, Whiteoak, \& Palmer 1983), NGC 7538(IRS1) (Palmer,
Gardner, \& Whiteoak 1984), and Sgr B2 (Gardner, Whiteoak, \& Palmer 1987).
Here we report observations of additional sources conducted in 1984
March--April.  Observations of the 6-cm OH lines have established that they
are closely associated with star formation, and that in many sources they
are the most highly variable of the OH maser lines.  Our primary interest
in this paper is in the accurate positions of the observed lines and the
variability of line intensities.  Subsequent papers in this series will
report a detailed study of both ground state and excited state OH and
H$_2$O maser emission in one region and a VLBA study of 4765-MHz emission
in several regions.

\section{Observations}

Observations were made with 25 antennas of the VLA.  For the 1984 March
observations, the VLA was in the CnB hybrid array; in April, the C-array.
These configurations provided beamwidths $\sim$4 arcsec.  For the 4765-MHz
observations, a correlator bandwidth of 0.781 MHz was divided into 512
spectral channels.  For the 4660- and 4750-MHz observations, a bandwidth of
3.125 MHz was used in conjunction with 256 channels.  On-line Hanning
smoothing of the resulting spectra and retention of every other channel
yielded 256- and 128-channel spectra with resolution of 3.0 kHz and 24 kHz
respectively.  Because of restricted data-taking rates at the VLA at that
time, observations were conducted in only one sense of polarisation (right
circular) and only the central 31 of the Hanning smoothed spectral channels
were recorded.  Both previous observations and theory had established that
these lines are unpolarised (\citealt{zppl}; \citealt{zp}).  The only
exception to date is MonR2, and in this case, it is argued that the
polarisation may arise from amplification of a polarised background source
(Smits, Cohen, \& Hutawarakorn 1998).  Because the 4765-MHz line is the
most commonly detected of the three OH transitions, we observed only this
line, except for W49A in which we observed all three OH lines. The velocity
resolution for all 4765-MHz observations was 0.19 km~s$^{-1}$ yielding a
total velocity coverage of $\pm$2.88 km~s$^{-1}$.  For the other two lines
observed in W49A, at 4750 MHz the resolution was 1.54 km~s$^{-1}$ yielding
a velocity coverage of $\pm$23.1 km~s$^{-1}$ and at 4660 MHz, 1.57
km~s$^{-1}$ yielding a velocity coverage of $\pm$23.6 km~s$^{-1}$.  While
the small velocity coverage at 4765 MHz was disappointing, we had access to
then recent unpublished observations of these lines with the 100-m
telescope at Effelsberg by Gardner (private communication, hereafter FFG)
to use to plan our observations.  The FWHM of the Effelsberg telescope at
this wavelength is $\sim$3 arcmin, while the FWHM of the primary beam of
the VLA antennas at this wavelength is $\sim$10 arcmin.  Therefore, any
source detected by FFG would also be detected by the VLA.  However, any
unexpected features outside the narrow velocity range that we were able to
cover were not recorded.  Other details for each source are provided in
Table~\ref{tab:tab1}.  All images produced from the observations extended
well beyond the FWHM of the primary beam because some positions were not
well known and to allow for serendipity.

In addition to data for each line channel, a pseudo-continuum ``channel 0''
was recorded.  These data consist of an average over the central 3/4 of the
spectral band.  For the 4765-MHz observations, channel 0 contained data
from 192 of the 3.0 kHz channels in the original (mostly unrecorded)
Hanning smoothed data set.  For the 4750- and 4660-MHz observations, the
pseudo-continuum was an average over 96 of the 24 kHz channels.  We checked
carefully that the resulting continuum images were not contaminated by line
emission by comparing them with images made from averages of known
line-free channels.  For DR21, channel 0 was not used. Rather, a normal VLA
continuum observation was made (50-MHz bandwidth), so line contamination is
not an issue.

The results for the 4765-MHz line observations are summarised in
Table~\ref{tab:tab2}.  To facilitate comparisons with recent work, both
B1950 and J2000 positions are given.  The position uncertainties are caused
by errors in positions of the phase calibrators, errors in array baselines,
and thermal noise.  Phase calibrator positions were adjusted to modern
values (uncertainties $<$0.01 arcsec) and led to a sensible shift only for
W49 (0.1 arcsec in declination).  The errors quoted are 4-$\sigma$ and are
dominated by thermal noise except for the highest signal-to-noise source,
DR21EX.  No attempt was made to de-convolve the linewidths for instrumental
broadening because in most cases the widths are close to our resolution.
In the following subsections, a brief discussion is given for each source.
Because the 4765-MHz emission is very variable, in Table~\ref{tab:tab3} we
summarise all flux density measurements since these lines were discovered
for the four detected sources.  In order to compare our results with those
obtained at different epochs by other observers when there was a
significant difference in pointing position, we have corrected their
measurements for the pointing offset from the VLA-determined positions by
assuming Gaussian antenna beams.  In a few cases the offsets are rather
large so that the Gaussian approximation is crude.  The flux densities
quoted are the corrected values, and this is indicated by a (C) following
the value in Table~\ref{tab:tab3}.  Note that in some cases, exact dates of
observations were not published.  We estimate that the errors in flux
densities in Table~\ref{tab:tab3} are dominated by the absolute calibration
uncertainties of the observations and are therefore about 10 per cent.

\subsection{S255}

This field is complex in the radio continuum; but, because the VLA in the
C-array resolves out structures larger than $\sim$30 arcsec, our continuum
image shows only the compact source G192.58-0.04 \citep{tt}.  This source
is located approximately between the optical nebulae S255 and S256, and
about 1 arcmin north of the cometary nebula PP56 which is associated with a
near-IR source, both H$_2$O and 1665-MHz ground state OH masers, and IRAS
06099+1800.  Although FFG had detected a maser feature of amplitude 1.2 Jy
at V$_{LSR}$= 6.61 km~s$^{-1}$ on September 15, 1983, approximately seven
months later we detected no emission exceeding 50 mJy (5 $\sigma$).  This
line has not been detected subsequently: 1 $\sigma$ upper limits of 230 mJy
and 780 mJy were obtained in 1989 June/July (Cohen, Masheder, \& Walker
1991, hereafter CMW) and in 1998 September 7 (Szymczak, Kus, \& Hrynek 2000,
hereafter SKH).  For a maser at the position found by FFG, these upper
limits should be increased by 16 per cent and 5 per cent, respectively.  At
the adopted distance to S255 and S256 (1.2 kpc; \citealt{israel}), the
projected separation between the 4765-MHz maser and PP56 corresponds to
$\sim$0.35 pc -- that is, the source of the 4765-MHz maser is well
separated from the other known masers in this field.

\subsection{K3-50}

The spectrum at the position of maximum maser intensity is shown in
Fig.~\ref{fig:fig1}a, the location of the maser on the
continuum image made from our data is shown in Fig.~\ref{fig:fig2}a.  The
position given in Table 2 is the position of the feature in the central
channel (V$_{LSR}$=-20.45 km~s$^{-1}$).  The maser is located on the edge
of the dominant compact HII region which was identified as the centre of an
ionised bipolar flow by
\citet{dgpr}.  [Our C-array data does not contain the short spacings necessary
to accurately image the extended flux density in this region which leads to
the negative bowl around the bright compact sources; a better image of all
of the radio continuum from this region at 6 cm is found in Roelfsema,
Goss, \& Geballe (1988).]  In Table~\ref{tab:tab3}, it can be seen that for
most of the 20 years since its discovery (\citealt{gmp1}, hereafter GMP), the
flux density, while variable, has stayed within a factor of two of 1 Jy --
except for the flaring episode observed by SKH.

K3-50 is part of the W58 complex of ionised and molecular gas (Dickel,
Goss, \& De Pree 2001).  Ground state OH is seen only in absorption in the
immediate vicinity of K3-50.  The nearest maser feature seems to be the
well known 1720-MHz source ON-3 which is 2.25 arcmin to the northeast which
corresponds to $\sim$5 pc projected separation at the 7.4 kpc distance of
the complex (\citealt{h75}, scaled to an 8.5 kpc distance to the Galactic
centre).  A more sensitive high resolution search for 18-cm masers would be
worthwhile.

\subsection{DR21EX}

DR21EX is located $\sim$2 arcmin north of the well known ground-state OH
source DR21(OH) which is located $\sim$3 arcmin north of the compact HII
region DR21 which dominates the continuum in this field.  The spectrum at
the maser intensity maximum is shown in Fig.~\ref{fig:fig1}b; the
location of the maser with respect to continuum is shown in
Fig.~\ref{fig:fig2}b.  The most striking thing about this maser is its
large separation from detectable continuum sources.  Table~\ref{tab:tab3}
traces the evolution of the feature near 5 km~s$^{-1}$ from 1968 to 2002.
In 1968, no feature was detected to a limit of $<$2 Jy (4-$\sigma$); in
1982 a single feature was discovered which reached an intensity of 12.8 Jy
in 1984.  Since then, the intensity has decreased irregularly so that in
2001, the flux density was $<$1 Jy and velocity features at $\sim$5
km~s$^{-1}$ were no longer the dominant features in the spectrum.

The flux densities in Table~\ref{tab:tab3} refer to components with
V$_{LSR}$ = 5.1--5.3 km~s$^{-1}$.  Features at other velocities (outside
the range observed in this paper) have been detected as noted in
Table~\ref{tab:tab3}.  Most of the features are located very close to the
position of the $\sim$5 km~s$^{-1}$ feature.  A notable exception is that
while Cohen, Masheder, \& Caswell (1995, hereafter CMC) found that the
position of the -4.0 km~s$^{-1}$ feature was the same as that of the 5
km~s$^{-1}$ feature; Palmer \& Goss (in preparation, hereafter PG) found a
feature at V$_{LSR}$=-3.70 km~s$^{-1}$ which was offset $\sim$15 arcsec
north of the 5 km~s$^{-1}$ position on 2001 March 02.

PG have detected a 1720-MHz maser within 1 arcsec of the 4765-MHz
5-km~s$^{-1}$ position; this is the only ground state OH maser from this
position in the literature (although many are reported near DR21(OH) and
DR21).  \citet{gd} report groups of H$_2$O-maser features both at the
DR21(OH) position and $\sim$1 arcmin north of the 4765-MHz feature.  PG
have detected another group of H$_2$O-maser features at the 4765-MHz
5-km~s$^{-1}$ position.

\subsection{W75N}

Ground state 1665-, 1667-, and 1720-MHz OH masers (as well as other masers)
are located very near this 4765-MHz maser.  The spectrum at the maser
maximum is shown in Fig.~\ref{fig:fig1}c; the maser location with respect
to the continuum sources in the field is shown in Fig.~\ref{fig:fig2}c.
There were two velocity components which have slightly different positions
(see Table 2).  The masers fall on the edge of a small HII region,
$\sim$1 arcsec from it's centre.  This offset corresponds to 2000 AU at the
accepted distance to this source, 2 kpc.

The ground state masers are scattered over a region about 5 arcsec in
diameter, but most are tightly clustered in a V-shaped region with sides
about 1.5 arcsec long (Hutawarakorn, Cohen, \& Brebner 2002).  A Zeeman
pair of 1720-MHz features was identified by those authors close to the
``point'' of the V.  The 4765-MHz feature is offset $\sim$1.6 arcsec west
of this pair (outside of the V).

The summary in Table~\ref{tab:tab3} shows that during the 20 years that
this source has been observed, its peak flux density has varied by more
than two orders of magnitude.

\subsection{W49A}

This is a complex region located $\sim$11.4 kpc distant (Gwinn, Moran, \&
Reid 1992).  The continuum shows at least 45 distinct HII regions (De Pree,
Mehringer, \& Goss 1997, hereafter DMG).  Many ground state OH masers in
all four lines have been detected in this complex region as well as
numerous H$_2$O masers.  All three 6-cm lines were observed, and they
revealed three principal sites of maser activity separated by $\geq$~1
arcmin.  One site shows only point-like 4765-MHz emission; another, only
point-like 4660-MHz emission; and the third, spatially extended 4750-MHz
and 4660-MHz emission with a broad velocity range as well as point-like
4765-MHz emission.  Comparisons of 18-cm OH masers in W49A with the 6-cm
masers are difficult for two reasons.  First, W49A contains many powerful
ground state masers in all four lines so that fainter features may be
confused.  Second, because W49A is $\geq$5 times further away than DR21 and
W75N for which we have noted closely spaced 4765-MHz and 1720-MHz features,
neither linear resolution nor sensitivity to intrinsically faint features
is comparable to those in the these sources.  The continuum image from our
data with the sites of 6-cm maser activity identified is shown in
Fig.~\ref{fig:fig2}d.  Note that our observations resolve out a significant
amount of the extended flux density.  We refer to the continuum components
shown in multi-configuration images of DMG both because they show the
extended flux density more accurately and because they have higher spatial
resolution.

\subsubsection{4765 MHz}

\begin{figure*}
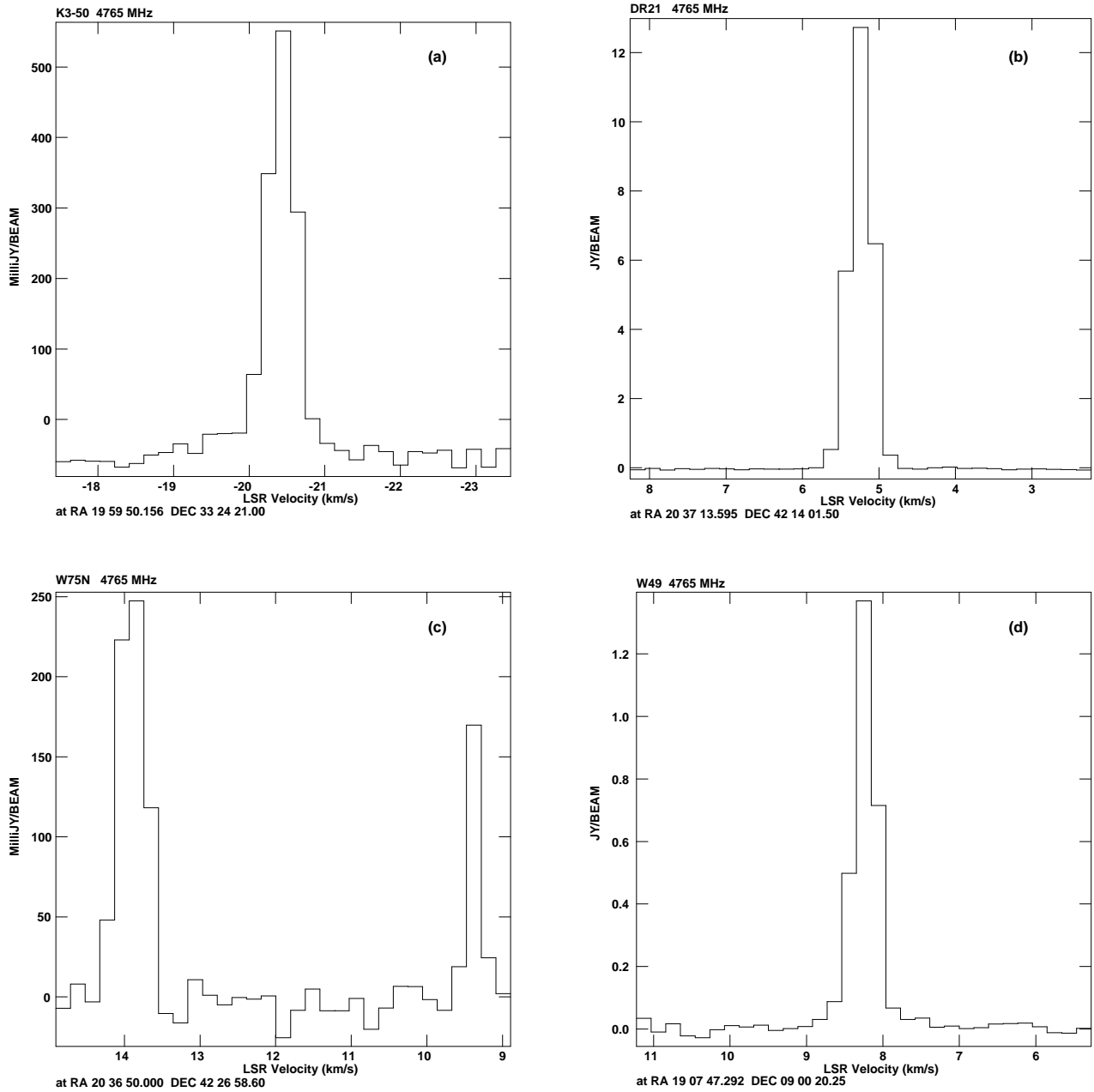


\centering
\begin{minipage}[c]{.45\textwidth}
   \centering
   \includegraphics[width=\textwidth]{fig1a.eps}
\end{minipage}
\hspace{0.05\textwidth}
\begin{minipage}[c]{.45\textwidth}
   \centering
   \includegraphics[width=\textwidth]{fig1b.eps}
\end{minipage}\\[4truemm]
\begin{minipage}[c]{.45\textwidth}
   \centering
   \includegraphics[width=\textwidth]{fig1c.eps}
\end{minipage}
\hspace{0.05\textwidth}
\begin{minipage}[c]{.45\textwidth}
   \centering
   \includegraphics[width=\textwidth]{fig1d.eps}
\end{minipage}
\caption{The 4765-MHz OH maser spectra obtained with the VLA.  The velocity 
resolutions are 0.19 km~s$^{-1}$.  (a) The spectrum toward K3-50.  (b) The
spectrum toward DR21EX.  (c) The spectrum toward W75N.  (d) The spectrum
toward W49A.  Spectra (a) -- (c) were obtained 1984 April 29; while (d) was
obtained 1984 March 23.  The rms noise levels and the synthesised beams for
each are given in Table 2.}
\label{fig:fig1}
\end{figure*}

\begin{figure*}
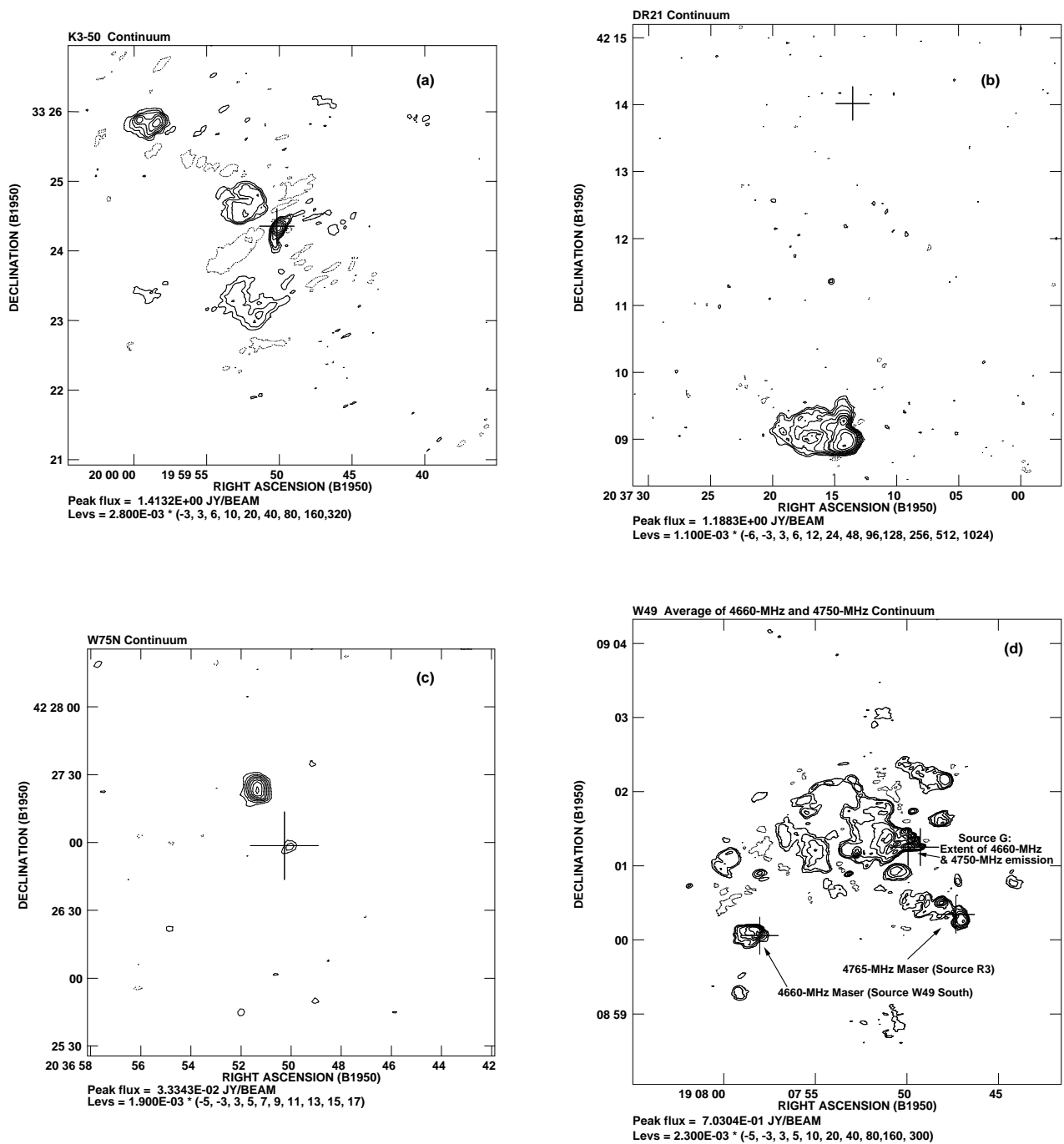

\centering
\begin{minipage}[c]{.45\textwidth}
   \centering
   \includegraphics[width=\textwidth]{fig2a.eps}
\end{minipage}
\hspace{0.05\textwidth}
\begin{minipage}[c]{.45\textwidth}
   \centering
   \includegraphics[width=\textwidth]{fig2b.eps}
\end{minipage}\\[4truemm]
\begin{minipage}[c]{.45\textwidth}
   \centering
   \includegraphics[width=\textwidth]{fig2c.eps}
\end{minipage}
\hspace{0.05\textwidth}
\begin{minipage}[c]{.45\textwidth}
   \centering
   \includegraphics[width=\textwidth]{fig2d.eps}
\end{minipage}\\[4truemm]

\caption{The 6-cm continuum emission observed with the VLA in this
experiment.  The contours are in units of the rms noise, beginning with
$\pm$3; dashed contours are negative.  For (a) -- (c), the synthesised
beams are $\sim$5 arcsec by $\sim$4 arcsec and crosses indicate the
4765-MHz maser position in each source.  (a) K3-50, showing significant
negative contours because of missing short spacings; (b) DR21 is the strong
source at the bottom, well separated from the maser (this observations was
made in the normal VLA continuum mode); (c) W75N; (d) The continuum
emission from the W49A region observed with the VLA on 1984 March 23.
Positions are indicated with crosses for the unresolved 4660-MHz maser near
W49 South and for the unresolved 4765-MHz maser near source R3. The extent
of 4660- and 4750-MHz emission near source G is indicated (see also figure
4).  The synthesised beam is 4.2 arcsec by 1.7 arcsec.}
\label{fig:fig2}
\end{figure*}

The spectrum at the position of maximum intensity is shown in
Fig.~\ref{fig:fig1}d.  This position is at the edge of the HII region
called R3 by DMG.  Table~\ref{tab:tab3} summarises the observations of the
8.2 and 8.6 km~s$^{-1}$ features since 1968.  Because the positions of the
two agree, and because before 1994.8 only the former was detected and since
1994.8 only the latter (except for the report by SKH in 1998), they are
treated as one in the table.  More than 6 years of closely spaced
monitoring data is provided by \citet[][hereafter S1, S2]{s97,s03}.  The
4765-MHz masers in W49A showed no dramatic flaring episodes, but a maser in
the 8.2 -- 8.6 km~s$^{-1}$ velocity range has been present with an
intensity varying on both long and short time-scales (but within a factor
of four of 1 Jy) for more that 30 years.

The presence of features at other velocities is noted in
Table~\ref{tab:tab3}.  The principal ones are those at $\sim$2.2 and
$\sim$11.9 km~s$^{-1}$.  Using the ATCA, \citet{de} determined that the
positions of both the 8.6 km~s$^{-1}$ and the 11.9 km~s$^{-1}$ features are
the same (and the same as that we determined for the feature shown in
Fig.~\ref{fig:fig1}d), while the position of the 2.2 km~s$^{-1}$ is offset
by 65 arcsec, north and east (into the Source G region, see below).
However, the ATCA positions for features in W49 have relatively large
errors in declination (2 -- 5 arcsec).  PG have improved the
positions of these features and strengthened these conclusions.

The 4765-MHz maser at 8.2/8.6 km~s$^{-1}$ is located within
$\sim$0.1 arcsec of several 1665-MHz and 1720-MHz features \citep{gm87}.
The 2.2 km~s$^{-1}$ position is in a very complex region with many ground
state masers.  The nearest are a collection of 1667-MHz masers offset by
$\sim$1 arcsec; the nearest 1720-MHz masers are $\sim$5 arcsec
distant.

\subsubsection{4660- and 4750-MHz Emission}
The 4750- and 4660-MHz lines are detected in a spatially extended source
located at the position of continuum source G (see Fig.~\ref{fig:fig2}d).
The 4660-MHz line is the more extended: covering Source G and extending to
the west across source B to approximately the position of source A (see
fig. 2 of DMG).  The 4750-MHz line has measurable intensity only in the
Source G region.  The 4660-MHz line also has a narrow velocity component
$\sim$3.5 arcmin away (projected separation: $\sim$11 pc) between the
main component of W49 South and the compact component W49 South-1 (source
names follow DMG).

\paragraph{Extended Component}

\begin{figure*}
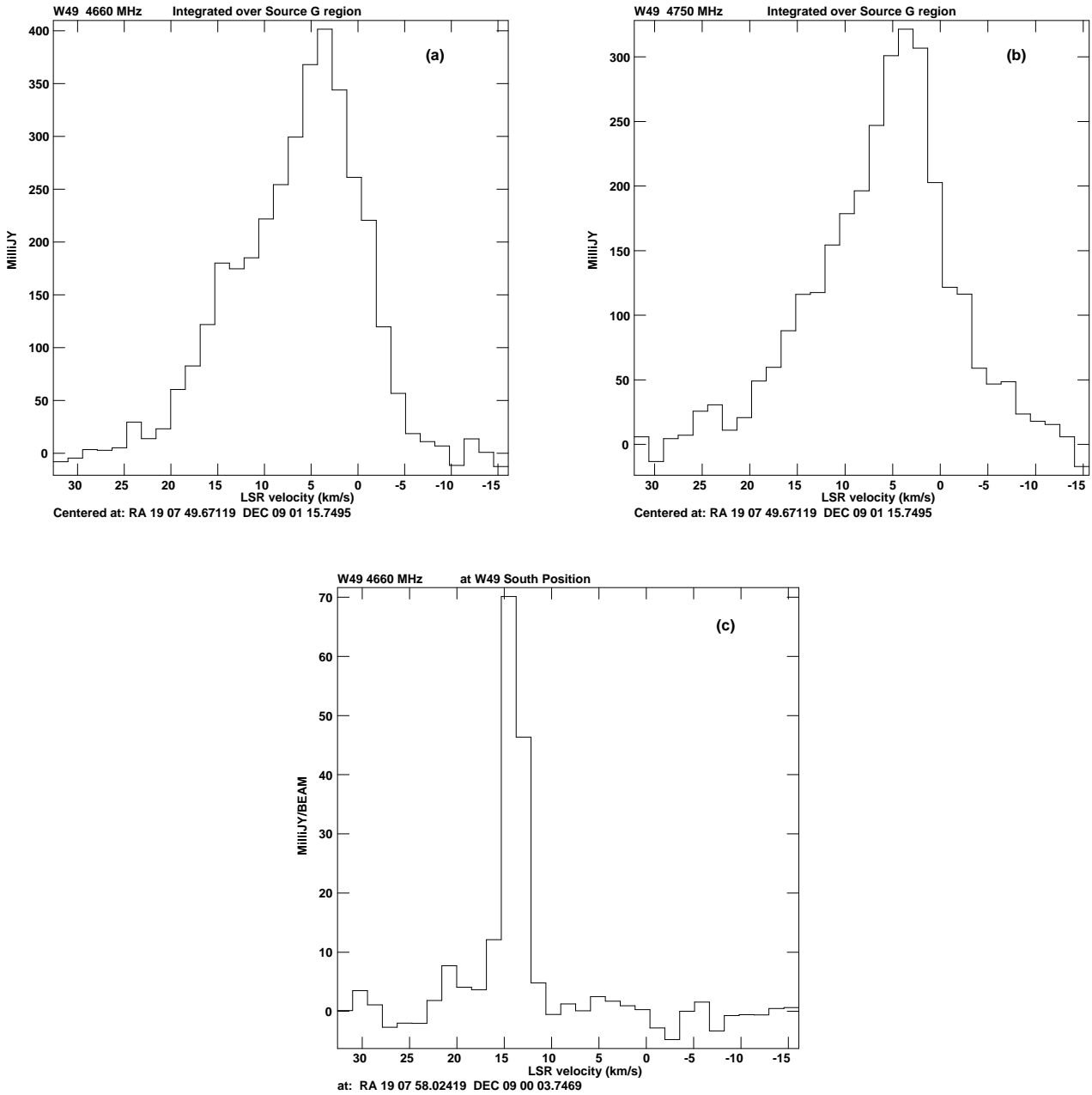

\centering
\begin{minipage}[c]{.45\textwidth}
   \centering
   \includegraphics[width=\textwidth]{fig3a.eps}
\end{minipage}
\hspace{0.05\textwidth}
\begin{minipage}[c]{.45\textwidth}
   \centering
   \includegraphics[width=\textwidth]{fig3b.eps}
\end{minipage}\\[4truemm]
\begin{minipage}[c]{.45\textwidth}
   \centering
   \includegraphics[width=\textwidth]{fig3c.eps}
\end{minipage}
\caption{The spectra of W49A at 4660-MHz and 4750-MHz observed with the VLA 
on 1984 March 23.  The velocity resolution is $\sim$1.5 km~s$^{-1}$, about
eight times broader than for the 4765-MHz masers in Figure 1. The
synthesised beams are $\sim$4.2 arcsec by $\sim$1.7 arcsec, and the rms
noise per channel is 3 mJy per synthesised beam.  (a) The spectrum of
4660-MHz emission integrated over 25 arcsec by 12 arcsec box including
sources G, B and A. (b) The spectrum of 4750-MHz emission in W49 integrated
over the same box as (a).  (c) The spectrum toward the 4660-MHz source near
W49 South.}
\label{fig:fig3}
\end{figure*}

The spectra of the extended components integrated over a box 25 arcsec x 12
arcsec are shown in Fig.~\ref{fig:fig3}a and b.  The integrated profiles of
the two lines are strikingly similar.  For both lines, the spatial extent
of the emission is at least 17 arcsec(RA) by 6 arcsec(dec).  However, when
inspected on a pixel-by-pixel basis, there are significant differences.
The 4660-MHz line clearly has two velocity maxima in much of this region.
The spectra toward individual positions are broad in velocity.  The FWHM of
the line ranges from $\sim$4 km~s$^{-1}$ to $\sim$17 km~s$^{-1}$.  In
contrast, the 4750-MHz line is dominated by a single velocity maximum
throughout the region.  The FWHM ranges from $\sim$4 km~s$^{-1}$ to
$\sim$18 km~s$^{-1}$.  The general agreement in profile shape and spatial
distribution of the two lines together with the differences on a
pixel-by-pixel basis, argue for excitation in the same volume of gas with
minor differences in excitation from point to point.

\begin{figure*}
\includegraphics[width=80mm]{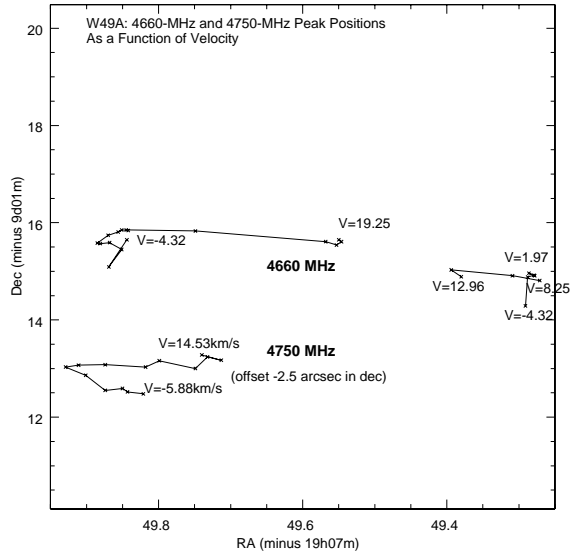}
\caption{The locations of 4660- and 4750-MHz emission peaks 
in the source G region of W49A.  The points are connected in order by
velocity. For the 4660-MHz line the positions plotted show both maxima
where two Gaussians could be fitted to the line.  Note that the velocity
changes systematically with position for the eastern peak.  The 4750-MHz data
is plotted 2.5 arcsec south of the actual declination.  The velocity of the
peak at 4750-MHz is derived from fits to single Gaussians.  Again, the
positional variation with velocity is rather systematic.}
\label{fig:fig4}
\end{figure*}

The positions of the peaks as a function of velocity is shown in
Fig.~\ref{fig:fig4}. For the 4660-MHz line, the position of the eastern
component (near source G) moves rather smoothly westward to the edge of
source B as the velocity increases from V$_{LSR}$=-4.32 km~s$^{-1}$ to
V$_{LSR}$=19.25 km~s$^{-1}$. The errors at each point depend on the signal
to noise; they are $\sim$0.05 arcsec near the line maximum and
$\sim$0.2 arcsec at each end.  The western component (near Source A) is
not visible at all velocities and is $<$6 arcsec.  Similarly, for the
4750-MHz line, the position of maximum intensity moves rather smoothly as a
function of velocity (mostly in RA) over about 4 arcsec.  The emission in
each channel is also slightly extended for both lines.

Note that these lines have much lower peak intensities than do the 4765-MHz
lines reported above and that they have much greater FWHM's (compare
Fig.'s 1 and 3).  Most of the published values must be read from
plots and are therefore not precise; but, including re-observations of
these lines with the VLA in 2001 August and November (PG), it
seems that neither of these lines has varied in the $\sim$30 years since
they were discovered.  The only difference between the VLA measurements and
single dish measurements is that single dish measurements of the 4660-MHz
line include varying amounts of the narrow feature near W49 South
(discussed in the next paragraph) which would add about 70 mJy at
$V_{LSR}$=$\sim$14 km~s$^{-1}$ if observed with a large single dish beam.
With the present angular resolution, it is not possible to determine
whether the emission is actually extended on this scale, or if it consists
of a clump of unresolved features.  However, PG have observed
both lines with the VLA with 0.4 arcsec resolution, and find that they do
not break up into collections of narrow features.

The region of the extended 4750- and 4660-MHz emission is the location of
many ground state masers in all four transitions [see fig. 7 of
\citet{gm87} which shows the positions superposed on a contour image of
Sources G, B, and A.]

\paragraph{Point-like component}
The spectrum at the maximum near W49 South is shown in
Fig.~\ref{fig:fig3}c.  The B1950 position is: 19$^h$07$^m$58.03$^s$, 09\degr
00\arcmin 03.4\arcsec.  This source resembles the 4765-MHz masers: it is
point-like (diameter $<$1.2 arcsec), and has a velocity FWHM $<$2.5 km
s$^{-1}$ (it may well be much narrower because our resolution is only
1.5~km~s$^{-1}$ for this line).  The velocity is $\sim$14.4 km~s$^{-1}$.
We have much less information about variability of this component because
it is not separated in the single dish measurements; on 2001 Aug 27, the
4660-MHz peak flux density was essentially the same as reported here
(PG). We have searched the entire 10 arcmin primary beam and no
other sources were detected at 4660 MHz.

At the position of the 4660-MHz maser, the upper limit for any
4750-MHz emission is $\sim$12 mJy ($<$20 per cent of the 4660-MHz emission).
The 4660-MHz maser is located $\sim$0.3 arcsec from
several 1612-MHz masers and within 0.5 arcsec of several 1667-MHz masers.
The nearest 1720-MHz masers are $\sim$1.1 arcsec away.  A possible
association of 4660-MHz emission with 1612-MHz emission has been noted by
\citet{wetal87} in SgrB2.

\section{Discussion}
Our primary goal was to determine precise positions for the 4765-MHz
masers.  The principal limitation is that our velocity coverage did not
allow us to include all now known velocity components.  As shown in Table
2, our position determinations are more than adequate to guide VLBI
observations.  We note that in cases where features at several velocities
are present, their separations are frequently not at the arc second scale
(like separations of 18-cm OH or H$_2$O maser spots) but by 10's of arc
second to arc minutes (more similar to the separations of clumps of 18-cm OH or
H$_2$O maser spots).  A byproduct of this work was a study of the
time-variability of these masers.  We have examined all published data, two
unpublished studies (FFG and RPZ), and some of our work in progress (PG).
As most other observers have discovered, the 4765-MHz masers are highly
time-variable.  Powerful flares such as reported for DR21EX by FFG and
this paper and, most spectacularly, in Mon R2 reported by \citet{sch} are
relatively rare.  Nevertheless, it is striking that over 20 -- 30 year
periods, three of the four maser sources detected in this paper have
exceeded 6 Jy at some time.  That is, although usually rather faint, these
masers occasionally reach flux density levels so that they could be
detected in an observation of a few minutes duration with a typical
25-meter telescope.  All known 4765-MHz features vary at some level, some
at the 10 per cent level and others by more than 100 per cent.

The region from which the spatially extended 4660-MHz and 4750-MHz emission
is detected in W49A is exceptional in several ways.  This region contains a
partial ring of HII regions [called source G by \citet{djww}] and several
other compact continuum components (see \citealt{dwgwm};
\citealt{wdwg}).  This region also contains the dynamical centre of an
exceptionally powerful H$_2$O maser outflow.  \citet{morris} found that the
velocity range exceeded 500 km~s$^{-1}$.  From a five epoch proper motion
VLBI study, \citep{gmr} located the dynamical centre of
the outflow within source G. The dynamical centre did not correspond to any
of the sub-components G1 -- G5 resolved by DMG with 0.8 arcsec resolution.
Subsequently G2, the source nearest the dynamical centre was resolved into
three sources: G2a -- G2c by \citet{dwgwm} with 0.04 arcsec resolution.
Recently, \citet{wdwg} in a study at 1.4-mm ($\sim$ 0.2 arcsec resolution)
identified an extension of G2b which corresponds to the dynamical centre of
the H$_2$O outflow.  A possible solution to the ``lifetime problem'' for
the large number of closely spaced compact HII regions (\citealt{dw};
\citealt{wc}) is that they may be confined by surrounding molecular clouds with
densities $n_{H_2} \sim 10^7$ -- $10^8$ cm$^{-3}$ (De Pree, Rodriguez \&
Goss 1995).  A density range of $\sim 10^7$ -- $10^8$ cm$^{-3}$ is required
to collisionally excite the $^2\Pi _{1/2}, J=1/2$ lines and it is above the
range of density required for excitation of the 1.4-mm methyl cyanide
emission observed by
\citet{wdwg}.  Therefore, the scenario of \citet{drg} would provide a
natural explanation for excitation of the observed OH and methyl cyanide
lines.  In summary, the entire region seems filled with dense gas in either
a neutral or an ionised state.

An important question is whether the spatially extended 4660-MHz and
4750-MHz emission is maser or thermal emission.  The peak flux density at
4750-MHz is 97 mJy/beam in a 4.11 arcsec x 1.72 arcsec beam,
corresponding to a brightness temperature $\sim$700 K.  For the 4660-MHz
line, the peak brightness is 168 mJy/beam in a 4.33 arcsec x 1.70 arcsec
beam corresponding to $\sim$1200 K.  Therefore it is hard to escape the
conclusion that the spatially extended 4660-MHz and 4750-MHz emission with
broad velocity widths near W49A source G is maser emission.

A picture that is compatible with the considerations above is that the 6-cm
OH lines are usually weakly inverted in the dense gas surrounding
star-forming regions.  Spatially extended, broad velocity, non-variable
4660-MHz and 4750-MHz lines are detectable whenever $\|\tau \|$ in the
dense regions is large enough.  Because the 4765-MHz lines are more
frequently found, conditions for large inversions must be reached
occasionally for these lines, although the column density and velocity
coherence reach high enough values to produce the narrow, intense lines at
4765-MHz along any line of sight only for relatively transient periods.

\section{acknowledgments}
We wish to express our appreciation of our colleague, Frank F. Gardner
(deceased), who worked with us in the early decades of this study.
P.P. thanks NRAO for hospitality during extensive visits while much
of this work was carried out.

\begin{table*}
\begin{minipage}{125mm}
\caption{Observing Log}
\label{tab:tab1}
\begin{tabular}{ccccccc} \hline
Source & Line & \multicolumn{2}{c}{Pointing Position} & Date of & V$_{lsr}$ 
& Synthesised Beam \\
  &(F$\rightarrow$ F) & RA(B1950) & Dec(B1950) & Observation & (km~s$^{-1}$) &
( arcsec x arcsec , \degr ) \\ \hline
S255 & 1$\rightarrow$0& 06 09 58.000 & 18 01 01.60 & 1984 Apr 29 & 6.61 &  6.24x4.22,PA=-66.2 \\
W49A & 1$\rightarrow$0 & 19 07 52.000 & 09 01 15.00 & 1984 Mar 23 & 8.25 & 4.26x1.83,PA=-79.6 \\
    & 1$\rightarrow$1 &  ''         &    ''       &   ''        & '' & 4.11x1.72,PA=-77.2 \\
    & 0$\rightarrow$1 &  ''         &    ''       &   ''        & '' & 4.33x1.70
,PA=-76.2 \\
K3-50 &1$\rightarrow$0& 19 59 48.000 & 33 24 00.00 & 1984 Apr 29 & -20.45 & 4.49x3.78,PA=-60.4 \\
DR21EX & 1$\rightarrow$0 & 20 37 14.000 & 42 12 00.00 & 1984 Apr 29 &5.24 & 4.25x4.13,PA=-51.1 \\
W75N  &1$\rightarrow$0& 20 36 50.000 & 42 26 55.60 & 1984 Apr 29 & 11.90
& 5.19x3.96,PA=-79.8 \\ \hline
\end{tabular}
\end{minipage}
\end{table*}

\begin{table*}
\begin{minipage}{125mm}
\caption{Results for 4765-MHz Line Observations}
\label{tab:tab2}
\begin{tabular}{ccccccccc} \hline
Source & RMS &\multicolumn{4}{c}{Maser Position} & Peak Flux & Velocity &
FWHM \\
         & (mJy) & RA(1950)  & Dec(B1950)  & RA(J2000) & Dec(J2000) & (mJy) &(km~s$^{-1}$) & (km~s$^{-1}$) \\ \hline
S255  &  10 &   --       &  --         &  --      &  --  & --    \\
W49A   &  15 & 19 07 47.32$\pm$0.01 & 09 00 20.6$\pm$0.1 & 19 10
10.95$\pm$0.01 & 09 05 17.7$\pm$0.1 & 1440$\pm$60 & 8.23$\pm$.01 &
0.35$\pm$.1 \\
K3-50 &  13 & 19 59 50.17$\pm$0.03 & 33 24 21.4$\pm$0.4 & 20 01
45.78$\pm$0.03 & 33 32 45.4$\pm$0.4 &600$\pm$40   & -20.43$\pm$.02  & 0.46$\pm$.1 \\
DR21EX & 17 & 20 37 13.55$\pm$0.01 & 42 14 01.2$\pm$0.1 &20 39
00.42$\pm$0.01 & 42 24 38.9$\pm$0.1 & 12830$\pm$200 & 5.23$\pm$.01 & 
0.37$\pm$.1 \\
W75N  & 8.5 &  20 36 50.01$\pm$0.05 &  42 26 58.7$\pm$0.6 & 20 38
36.45$\pm$0.05 & 42 37 35.1$\pm$0.6 & 280$\pm$30 & 13.90$\pm$.05 & 
0.42$\pm$.1 \\  
       &     &  20 36 49.99$\pm$0.08 &  42 26 58.3$\pm$0.9 &20 38
36.43$\pm$0.08 & 42 37 34.7$\pm$0.9  & 170$\pm$30 & 9.37$\pm$.10 &
0.23$\pm$.1 \\ \hline
\end{tabular}
\end{minipage}
\end{table*}

\begin{table*}
\begin{minipage}{125mm}
\caption{Summary of 4765-MHz Emission Variability}
\label{tab:tab3}
\begin{tabular}{lllll} \hline
Source & Date & S$_{max}$ & Reference & Comments \\
       &      &  (Jy)      &           &  \\ \hline
K3-50 & 1982 (early) & 0.6  &  GMP & \\
      & 1983 Aug 23 & 0.6  &  FFG & \\
      & 1984 Apr 29 & 0.6  & (this paper) & \\
      & 1991        & 1.0  & CMC & \\
      & 1998 Jun 30 & 6.8  & SKH & \\
      & 1998 Oct 20 & 2.4  & SKH & \\
      & 1998 Dec 1  & 2.2  & SKH & \\
      & 2001 Aug 9  & 1.9  & PG & \\
      & 2002 Jan 10 & 2.3  & PG & \\
DR21EX & 1968 Feb & $<$2.0(C) & \citet{zppl} & \\
      & 1982 (early) & 4.9(C) & GMP & \\
      & 1983 Aug 03 & 10.9 & FFG & \\
      & 1984 Apr 29 & 12.8 & (this paper) & \\
      & 1989 Jun/Jul & 7.7(C) & CMW & also V$_{lsr}$=-4.0 km~s$^{-1}$\\
      & 1991       & 3.8  & CMC &  also V$_{lsr}$=-3.2 km~s$^{-1}$ (no -4.0
km~s$^{-1}$) \\
      & (1994.8 -- 1996.4) & $\sim$2. & S1; S2 & observed 27 times in
interval \\
       &  &   & &  also V$_{lsr}$=3.6, 4.9, and 5.5 km~s$^{-1}$ \\
      & 1998 Jun 30 & 6.4(C) & SKH & also V$_{lsr}$=-4.0 km~s$^{-1}$ \\
      & 2001 Mar 02 & 1.0 & PG & also V$_{lsr}$=-3.7 km~s$^{-1}$ \\
      & 2001 Jun 12 & 1.1 & PG & \\
      & 2001 Nov 15 & 0.8 & PG & V$_{lsr}$=$\sim$5 km~s$^{-1}$ no longer
dominant feature \\
      &  &   & & at least four features: V$_{lsr}$=2.28 -- 5.71 km~s$^{-1}$ 
\\
W75N & 1983 Sep 15 & 0.1, 0.2 & FFG & V$_{lsr}$= 13.90, 9.37 km~s$^{-1}$ \\
     & 1984 Apr 29 & 0.3, 0.2 & (this paper) &  V$_{lsr}$= 13.90, 9.37
km~s$^{-1}$ \\
     & 1989 Jun/Jul & $<$0.3(C) & CMW & \\
     & 1998 Aug 14 & 2.9 & SKH & V$_{lsr}$=10.4 km~s$^{-1}$  \\
     & 1998 Nov 20 & 4.3 & SKH & V$_{lsr}$=10.3 km~s$^{-1}$ \\
     & 1998 Dec 2 & 6.1 & SKH & V$_{lsr}$= 10.4 km~s$^{-1}$ \\
     & 1998 Dec 19 & 10.0 & SKH & V$_{lsr}$= 10.3 km~s$^{-1}$ \\
     & 2001 Mar 02 & 0.04 & PG & V$_{lsr}$=10.49 km~s$^{-1}$   \\
W49 & 1968 Feb & 1.0 & \citet{zppl} & probably inadequately resolved \\
    & 1969 (early) & 2.0 & \citet{zp} & \\
    & 1974 Feb & 2.5(C) & RPZ\footnote{RPZ=Rickard, Palmer, \& Zuckerman (unpublished
manuscript)} & \\
    & 1980 Sep & 2.0 & JSS\footnote{JSS=Jewell, Schenewerk, \& Snyder 1985} & \\
    & 1981 Aug & 1.6 & JSS & \\
    & 1982 (early) & 1.3 & GMP & \\
    & 1982 May & 1.6 & JSS &  \\
    & 1983 Aug 23 & 1.4 & FFG & \\
    & 1984 Mar 23 & 1.4 & (this paper) & \\
    & 1989 Jun/Jul & 1.6(C) & CMW & \\
    & 1991    & 2.5 & CMC  & also V$_{lsr}$=2.2 km~s$^{-1}$\\
    & (1994.8 -- 1998.3) & $\sim$2. & S1; S2 & 
observed 80 times in this interval, V$_{lsr}$= 8.6 km~s$^{-1}$ \\
   & & & & also V$_{lsr}$= 2.2, 3.62, and 8.36 km~s$^{-1}$ \\
    & 1998 Jul 05 & 3.5  & SKH & V$_{lsr}$= 8.2 km~s$^{-1}$ \\
    & 2000 Sep 15/16 & 0.2 & \citet{de} &  V$_{lsr}$= 8.6 km~s$^{-1}$ \\
    &  &  &  & also V$_{lsr}$= 2.2, 2.7, and 11.9 km~s$^{-1}$ \\
    & 2001 Aug 09 & 0.3 & PG & V$_{lsr}$= 8.61 km~s$^{-1}$ \\
    &  &  &  & also V$_{lsr}$= 2.2 and 11.9 km~s$^{-1}$ \\ \hline
\end{tabular}
\end{minipage}
\end{table*}

\end{document}